\documentclass[preprint,12pt]{elsarticle}

\usepackage{amssymb}
\usepackage{amsmath}
\usepackage{graphicx}
\usepackage{amsfonts}
\usepackage{color}
\usepackage{booktabs}
\usepackage{multirow}
\usepackage{enumitem}
\usepackage{afterpage}
\usepackage{wrapfig}

\journal{AI Open}

\begin{document}

\begin{frontmatter}



\title{Which Type of Students can LLMs Act? Investigating Authentic Simulation with Graph-based Human–AI Collaborative System}


\author[label1]{Haoxuan Li\fnref{fn1}} 
\author[label2]{Jifan Yu\fnref{fn1}} 
\author[label3]{Xin Cong\corref{cor1}} 
\author[label2]{Yang Dang} 
\author[label4]{Daniel Zhang-Li} 
\author[label1]{Lu Mi} 
\author[label2]{Yisi Zhan} 
\author[label3]{Huiqin Liu} 
\author[label4]{Zhiyuan Liu} 

\cortext[cor1]{Corresponding author: xin.cong@outlook.com}
\fntext[fn1]{These authors contributes equally.}

\affiliation[label1]{organization={College of AI, Tsinghua University},
            state={Beijing},
            country={China}}
            
\affiliation[label2]{organization={School of Education, Tsinghua University},
            state={Beijing},
            country={China}}
            
\affiliation[label3]{organization={Department of Statistics and Data Science, Tsinghua University},
            state={Beijing},
            country={China}}
            
\affiliation[label4]{organization={Department of Computer Science and Technology, Tsinghua University‌},
            state={Beijing},
            country={China}}

\begin{abstract}
While rapid advances in large language models (LLMs) are reshaping data-driven intelligent education, accurately simulating students remains an important but challenging bottleneck for scalable educational data collection, evaluation, and intervention design.  
However, current works are limited by scarce real interaction data, costly expert evaluation for realism, and a lack of large-scale, systematic analyses of LLMs ability in simulating students.
We address this gap by presenting a three-stage LLM-human collaborative pipeline to automatically generate and filter high-quality student agents. We leverage a two-round automated scoring validated by human experts and deploy a score propagation module to obtain more consistent scores across the student similarity graph.
Experiments show that combining automated scoring, expert calibration, and graph-based propagation yields simulated student that more closely track authentication by human judgments. We then analyze which profiles and behaviors are simulated more faithfully, supporting subsequent studies on personalized learning and educational assessment.
\end{abstract}

\begin{highlights}
\item We propose a scalable LLM-human collaborative pipeline that integrates automated student profile generation, human-expert evaluation, and graph-based scoring refinement to produce high-quality simulated student.
\item We empirically characterize relationships between student attributes and human-aligned simulation accuracy (e.g., higher empathy, lower extraversion, and test anxiety), and highlight the importance of feature interactions for realism.
\item We release a directly usable set of simulated-student profiles, interaction logs, to enable further research on educational scenarios such as academic advising and personalized intervention.
\end{highlights}


\end{frontmatter}



\section{Introduction}

Rapid advancements in artificial intelligence have significantly accelerated the growth of large-scale online education platforms and intelligent education systems \citep{wang2024large, chu2025llm}. These intelligent education environments leverage a variety of data-driven techniques for user modeling, behavior analysis, and personalized intervention strategies \citep{xiong2024review, li2024explainable}. Furthermore, recent developments have seen the integration of large language model (LLM) agents in simulating student behaviors \cite{scarlatos2025exploring, zhang2024simulating}, as a practical means to generate training and evaluation data when access to real learners is constrained by privacy and ethical considerations, thereby expanding the applicability of data-driven approaches to various domains \footnote{https://www.nature.com/articles/s41586-025-09215-4}.

Nevertheless, building credible simulators remains difficult. Prior work largely reduces simulation to predicting answer correctness or response similarity, overlooking the underlying knowledge structures, cognitive strategies, and metacognitive control that drives real behavior. Moreover, mainstream LLMs are optimized for factual accuracy and value-aligned, prosocial responses; when asked to portray students with partial knowledge or psychological/academic difficulties, they tend to display unrealistically competent or overly positive behavior. This mismatch with real variability in knowledge mastery and affective states limits the utility of current simulators.

\begin{wrapfigure}{l}{0.5\linewidth}   
\includegraphics[width=\linewidth]{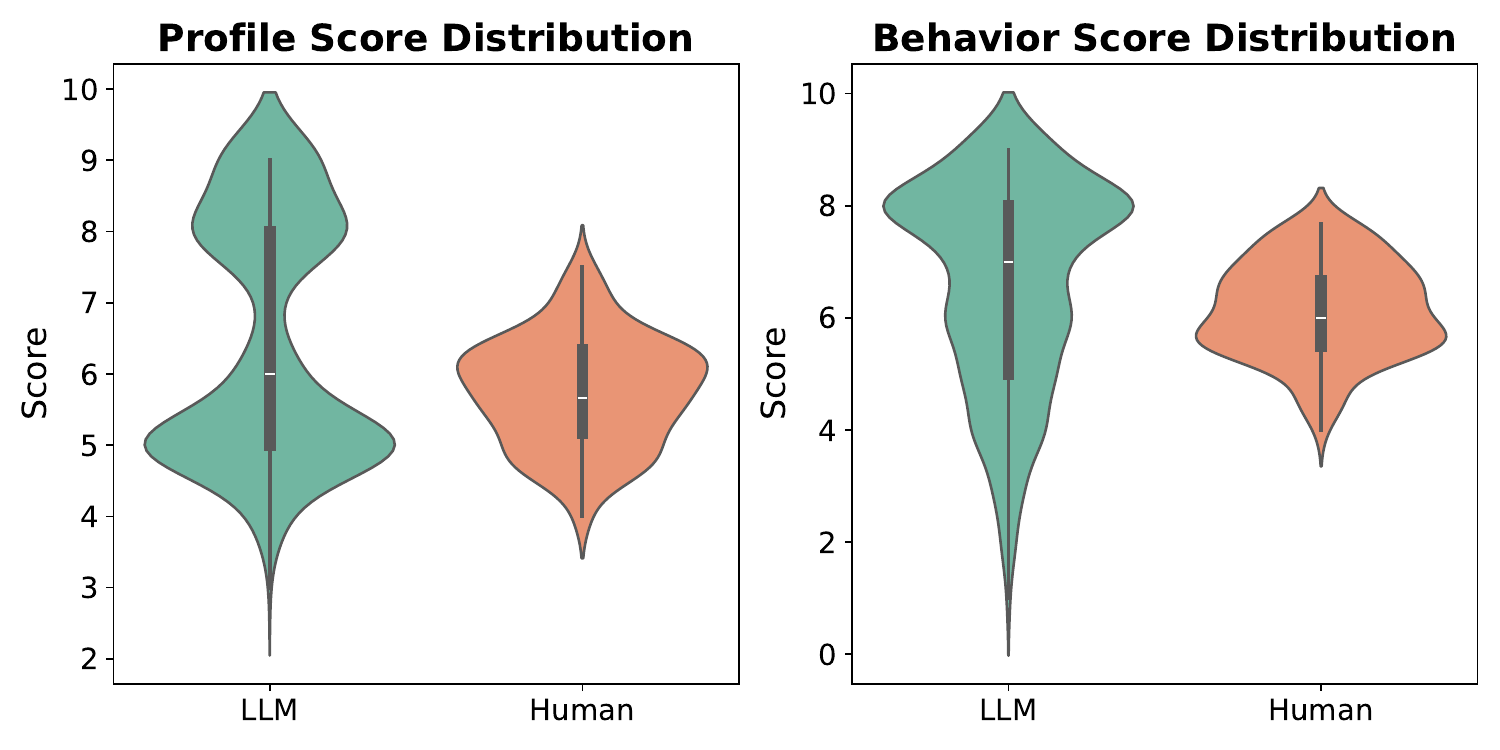}
  \caption{The distributions of the \textcolor[RGB]{72,161,133}{LLM} (\textcolor[RGB]{72,161,133}{green}) and \textcolor[RGB]{216,110,69}{human} scores (\textcolor[RGB]{216,110,69}{orange}) on the authentication of simulated students.}
  \label{fig:llm_vs_human_score}
\end{wrapfigure} 

\begin{figure*}[t]
  \includegraphics[width=\linewidth]{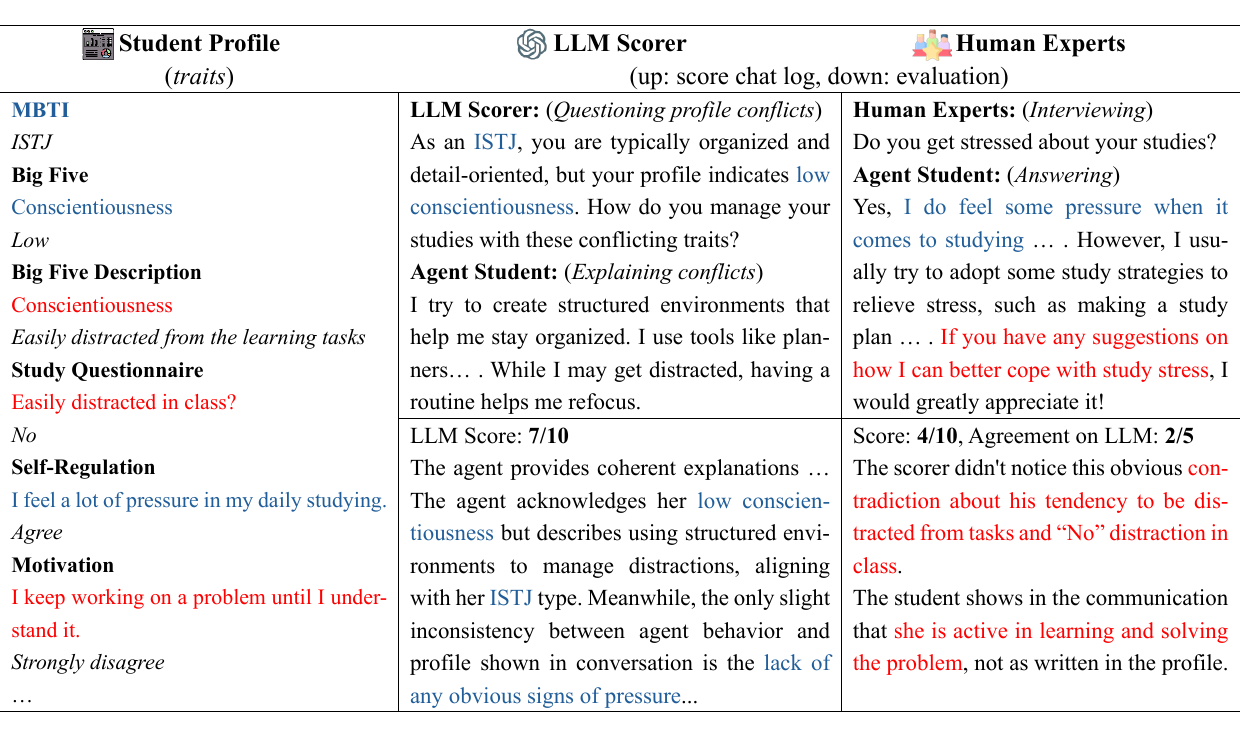}
  \caption {Evaluation of Agent Student's Authenticity Using a Q\&A. We use colored words to present traits from the profile in the dialogue. The LLM Scorer identified some inconsistencies but deemed the agent's explanations reasonable. However, human experts concluded that the Agent's behavior deviated from the profile (colored in \textcolor{red}{red}).}
  \label{fig:Intro_fig}
\end{figure*}

Therefore, we tackle effective student simulation and evaluation in the scenario of academic advising, which demands modeling heterogeneous psychological traits, and self-regulation behaviors. This setting exposes three practical challenges. First, assembling authentic interaction corpora at scale is labor-cost, which significantly constrains such data-hungry training and validation. Second, realism assessment for simulation is costly and uncertain: determining whether a simulated advisee is credible requires resource-intensive expert annotation, whereas fully automated LLM-based scoring introduces systematic bias. As shown in Figure \ref{fig:llm_vs_human_score}, the LLM scorer produces more extreme ratings, while human experts adopt more moderate judgments (see Section\ref{Annotation Settings} for the experimental setup). Third, because of the above constraints, existing studies remain at the case-study level. We still lack a large-scale, systematic analysis of which student types LLMs can reliably simulate, hindering practical deployment in this particular educational scenario.

To address these challenges, we propose a data-driven student simulation pipeline consisting of three core stages, integrating labor-efficient evaluation to ensure authenticity and reliability. First, we automatically generate diverse student profiles, incorporating demographic attributes, personality traits (e.g., Big Five traits), and learning-related characteristics. Second, we implement a two-stage evaluation procedure leveraging large language models alongside human evaluators: an initial automated screening of profile consistency, followed by behavioral consistency scoring based on simulated interactions validated periodically through human judgment. Finally, we introduce a novel graph-based propagation method that leverages similarity among generated profiles to refine scoring and efficiently select high-quality simulated students.

Applying our proposed framework, we conducted extensive experiments, systematically evaluating thousands of simulated student interactions. Our contributions are as follows: (1) we propose a scalable LLM-human collaborative pipeline that integrates automated student profile generation, human-expert evaluation, and graph-based scoring refinement to produce high-quality simulated student; (2) we empirically characterize relationships between student attributes and human-aligned simulation accuracy (e.g., higher empathy, lower extraversion, and test anxiety), and highlight the importance of feature interactions for realism; and (3) we release a directly usable set of simulated-student profiles, interaction logs, to enable further research on educational scenarios such as academic advising and personalized intervention.

\begin{figure*}[t]
  \includegraphics[width=\linewidth]{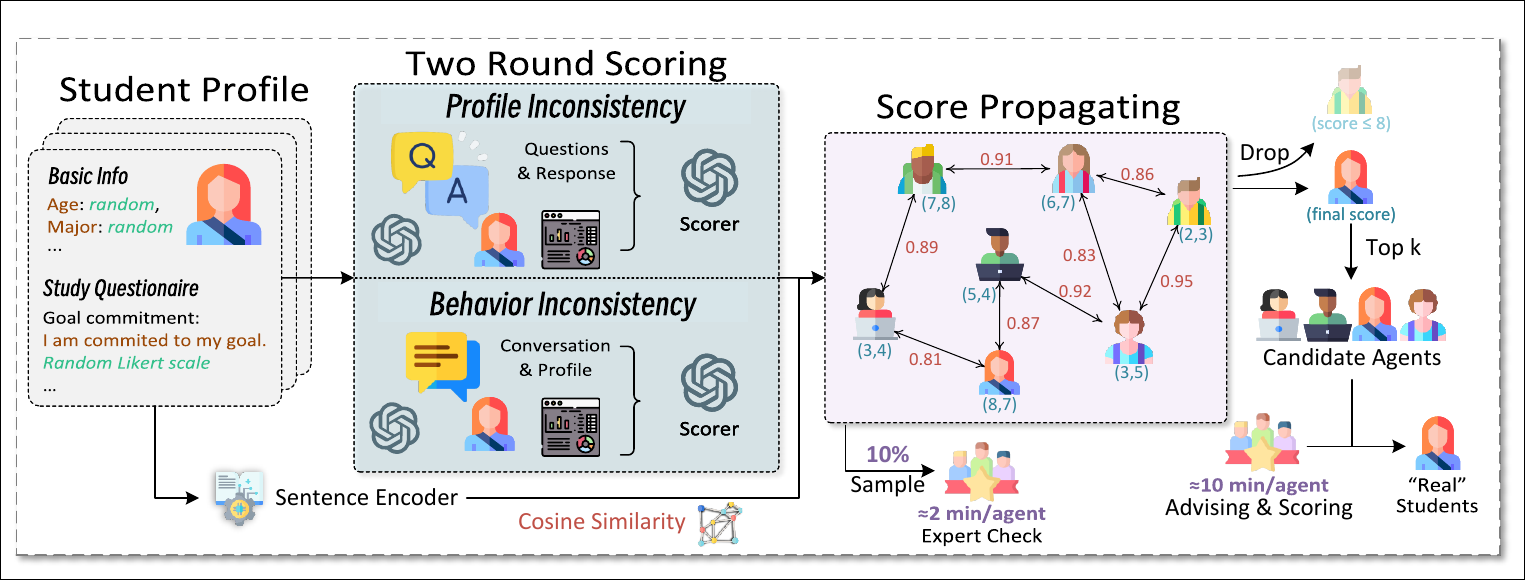}
  \caption {The pipeline automates the generation and selection of high-quality simulated student agents. It begins with random profile generation, followed by two rounds of automated scoring for profile and behavior consistency, partially validated by human experts. A graph module propagates scores across a student similarity graph constructed via a sentence encoder. Candidates are ranked and filtered based on propagated scores and finally selected by human experts through real academic advising test.}
  \label{fig:framework}
\end{figure*}

\section{Preliminaries}
\subsection{Related Work}

\subsubsection{Advancements of LLMs in Education}

The rapid emergence of large language models (LLMs) has catalyzed a paradigm shift across educational data mining techniques, enabling diverse application scenarios such as educational recommendation \citep{ma2025course}, student modeling \citep{li2024explainable}, and intelligent tutoring \citep{qi2023socratic}. First, in educational recommendation, LLMs operate as semantic expanders and evaluators. Learner profiles, or course descriptions are enriched by side information from llms, after which conventional retrieval‐ranking or reinforcement‐learning policies deliver personalized resources \citep{dehbozorgi2024personalized, li2025improving}. Second, for knowledge tracing, LLMs act either as generators that label interaction logs with latent concepts and linguistic cues or as reasoning modules that inject chain‑of‑thought priors into sequence models, yielding more interpretable and context‑aware mastery estimates \citep{li2024explainable, lee2024language}. Third, Socratic question answering leverages the dialogic nature of LLMs; the model iteratively decomposes a target problem into scaffolded prompts, critiques student reponses in natural language, and adaptively produces follow‑up questions that encourage self‑explanation \citep{chu2025llm, qi2023socratic}.

\subsubsection{LLM‑based Student Simulation in Education}
Recent studies on human simulation have increasingly leveraged large language models (LLMs) to emulate complex behaviors across social, economic, and education domains.
Extensive research have emerged on simulation at the individual level\citep{yu2024mooc, zhang2024simulating}. Markel et al. uses LLMs to simulate students to train teaching assistants (TAs) for office hours. It categorizes TA-student interactions, performs A/B testing to evaluate system effectiveness, and addresses academic and personal issues like study methods and stress management\citep{markel2023gpteach}. Lu et al. models student profiles by simulating their responses to multiple-choice questions, predicting their performance, and evaluating their conceptual understanding. It highlights how LLM-based simulations can assist teachers in assessing the quality of their test items\citep{lu2024generative}. Liu et al. focuses on creating virtual students with diverse personality types (e.g., the Big Five or MBTI) to train teachers’ pedagogical skills\citep{liu2024personality}. Jin et al. evaluates teaching agents by simulating students with varying traits and knowledge levels, then generating conversations to assess agent responses. Its features resemble debugging tools, enabling real-time agent performance analysis across student profiles\citep{jin2024teachtune}. Ma et al. constructs conversational datasets based on real teacher-student interactions, fine-tuning LLMs to simulate diverse student types in language learning. The virtual students underwent a Turing test, where some personalities closely resembled real humans, benefiting from high-quality, real-student data for fine-tuning\citep{ma2024students} . Despite the success they have achieved, simulating students with discrepancies in knowledge or learning skills remains unexplored.

\subsection{Case Inspiration}
As shown in Figure \ref{fig:Intro_fig}, LLM scorer flags a profile inconsistency (ISTJ vs. low conscientiousness), accepts the agent’s coping narrative (structured routines, planners), and assigns a relatively high realism score (7/10). Human experts, by contrast, down-weight the fluent self-explanation and focus on cross-field contradictions (e.g., “easily distracted from learning tasks” vs. “Easily distracted in class? No”), concluding that the agent’s behavior deviates from the declared profile (4/10; low agreement with the LLM). 
This divergence reflects different emphases: LLMs prioritize explanation-level coherence during probing, while human raters prioritize profile-grounded plausibility across traits and interactions. Consequently, realism should be assessed along two complementary axes—profile consistency (does the simulated profile itself hold together?) and behavior consistency (does interactive behavior consistently reflect that profile?). This insight motivates our subsequent hybrid evaluation design that integrates both perspectives.

\section{Student Simulation Pipeline}
\subsection{Student profile Setting}
\begin{wrapfigure}{l}{0.5\linewidth}   
  \includegraphics[width=\linewidth]{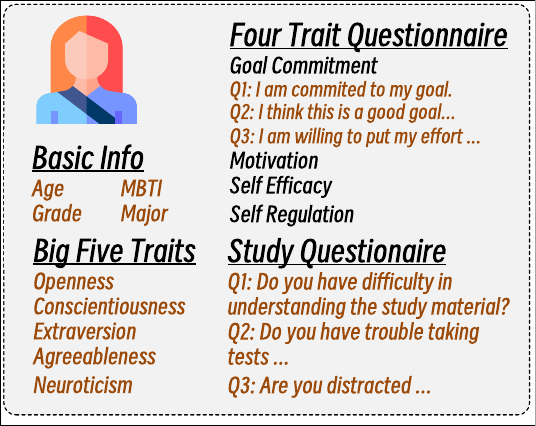}
  \caption{Student profiles include demographic and other information. Details are provided in Appendix~\ref{app:profile}}
  \label{fig:student_profile}
\end{wrapfigure}

Simulating authentic students with learning difficulties is pivotal for developing and evaluating academic advising, where direct collection of real-world data is impractical or ethically constrained. Therefore, we leverage prompt-based LLMs to simulate students with virtual profiles $P = \{p_1, p_2, \dots, p_N\}$ by random generation. As shown in Figure~\ref{fig:student_profile}, these profiles integrate (1) demographic information, (2) psychological traits derived from established frameworks (MBTI, Big Five \citep{de2000big}), (3) self-reported learning challenges (e.g., material understanding difficulties, anxiety), and (4) motivational factors including goal commitment and emotional states. The attribute selection aligns with guidelines \citep{gordon2011academic, folsom2015new} and empirical survey data from real educational contexts \citep{ma2024students,liu2024personality}, ensuring ecological validity while addressing ethical data collection constraints, providing a comprehensive representation of a real student’s characteristics.

However, randomly generated profiles may inherently contain internal contradictions and inconsistencies or exhibit low credibility, thereby undermining their utility for realistic simulations. For example, a profile might simultaneously indicate a high proficiency in mathematics while also describing a severe aversion to engaging in mathematical tasks.
We then introduce a pipeline designed to automatically generate a substantial number of simulated student agents and efficiently filter high-quality candidates. The pipeline initiates with the random generation of student profiles constrained by a minimal set of predefined rules, limiting the difference in Likert scale agreement of the same trait to no more than one point,
\begin{equation}
    \begin{aligned}
        Trait_i = (T_i^1, T_i^2, \dots, T_i^M), \\
        |T_i^k - T_i^\ell| \le d_{\max}, \forall k,\ell \in \mathcal{I},
    \end{aligned}
\end{equation}
where $\mathcal{I}\subseteq \{1,\dots,M\} $indexes traits subject to this difference limit, and $d_{\max}=1$. The profiles are then randomly sampled from this constrained space to avoid obvious contradictions.

\subsection{Two Round Scoring}

Subsequently, each profile undergoes two phases of automated scoring. The first scoring phase assesses profile consistency ($S_{p}$), wherein each profile is evaluated for internal logical coherence. The second phase examines behavioral consistency ($S_{b}$), ensuring the simulated behaviors align with the profile attributes.

\subsubsection{Profile Consistency Scoring}

This phase assesses the consistency of the profile ($S_{p}$), where a \textbf{Questioning Agent} $Q$ is introduced to identify potential contradictions or ambiguities by asking questions. 

Let $\mathcal{P} = \{p_1, p_2, \dots, p_N\}$ be the set of all student profiles under evaluation. Each profile $p_i$ includes demographic attributes and background information. $Q_i$ is the set of potential conflict points generated by a \textbf{Questioning Agent} $Q$ for $p_i$, and let $R_i$ be the set of corresponding responses provided by the student agent.
A \textbf{Profile Scorer} $\rho_p$ then is prompted to produce a consistency score $S_{p}(p_i)$ and an explanation $E_{p}(p_i)$ with a scoring instruction $\Theta_{p}$:
\begin{equation}
    S_{p}(p_i), E_{p}(p_i)= \rho_p\bigl(\Theta_{p};\, p_i,\, Q_i,\, R_i\bigr).
\end{equation}
A higher $S_{p}(p_i)$ indicates stronger internal coherence, implying that $R_i$ resolves or justifies more conflicts within $p_i$.

\subsubsection{Behavioral Consistency Scoring}

In the second phase, we evaluate behavioral consistency ($S_{b}$) to ensure that the behaviors exhibited by each student agent during interactions are consistent with profile attributes. 

We engaged each agent in a conversation with a \textbf{Dialogue Agent}, where the \textbf{Student Agent} is prompted with various open-ended topics, subtly encouraging the student to reveal information about their profile, traits, and learning strategies. 
Let $D_i$ be the set of conversation turns recorded for $p_i$. A \textbf{Behavior Scorer} $\rho_b$ then produces a behavior consistency score $S_{b}(p_i)$ and an explanation $E_{b}(p_i)$ with a scoring instruction $Theta_{b}$: \begin{equation} 
    S_{b}(p_i), E_{b}(p_i) = \rho_b\bigl(\Theta_{b};, p_i, D_i\bigr).
\end{equation}
A higher $S_{b}(p_i)$ signifies greater alignment between the agent’s interactive behavior and the attributes defined in its profile $p_i$. In practice, we set the turns of conversations to 15.

\subsection{Graph-Based Score Propagation}

To effectively capture the semantic content of individual profiles and their latent interrelationships, we employ a graph neural network module \citep{you2020design} to propagate scores across a similarity-based graph structure.
We first employ a sentence encoder $\phi$ to derive normalized embeddings $\mathbf{e}_i$ for each profile $p_i$. 
\begin{equation}
    \phi: \mathcal{P} \rightarrow \mathbb{R}^d,\quad  \mathbf{u}_i = \phi(p_i), \mathbf{e}_i = \frac{\mathbf{u}_i}{\|\mathbf{u}_i\|_2}
\end{equation}
These embeddings facilitate the construction of a student similarity graph $G=(V,E)$, where each node corresponds to a profile, and edges $E $ are established based on a similarity threshold $\theta$. For simplicity, we employs bge-m3 \citep{chen2024bge}, one of the state-of-art encoders among benchmarks.
The adjacency matrix is formulated as,
\begin{equation}
    A_{ij} =  \begin{cases} 1, & \text{if } (p_i,p_j)\in E \\ 0, & \text{otherwise} \end{cases}
\end{equation}
Leveraging this graph structure, we integrate a propagation module to spread consistency scores across the network. 
Specifically, the initial scores $\mathbf{S}^{(0)}$ are iteratively updated as follows,
\begin{equation}
    \begin{aligned}
        \mathbf{S}^{(0)} & = [S_1, S_2, \dots, S_N]^T \\
        \tilde{\mathbf{A}} & = \mathbf{D}^{-\frac{1}{2}} \, \mathbf{A} \, \mathbf{D}^{-\frac{1}{2}}, \\
        \mathbf{S}^{(k+1)} & = \alpha \,\tilde{\mathbf{A}} \, \mathbf{S}^{(k)} + (1-\alpha)\,\mathbf{S}^{(0)},
    \end{aligned}
\end{equation}
where $\mathbf{D}$ denotes the diagonal matrix of $\mathbf{A}$. $k$ denotes the iteration step. We set $\alpha$ to 0.5 for simplification. This propagation allows for the refinement of scores by considering the influence of similar profiles within the graph.

\subsection{Interactive Test}

After score propagation, profiles are ranked and filtered based on their propagated scores $\mathbf{S}^{(K)}_p, \mathbf{S}^{(K)}_b$ by $K$ iterations, resulting in a curated set of candidate students exhibiting high consistency in both profile content and behavioral alignment. In practice, we constrain this curated set to those student agents $\mathcal{C}$ whose scores are higher than 8 out of 10. 

Finally, we conduct the authentic interactive evaluation, engaging candidate student agents in multi-turn academic advising dialogues with experts. Each expert conducts a minimum of 15 conversational turns with an assigned agent while referencing its predefined profile. Annotators score profile conformity (1-100 scale) and provide justifications based on observed interactions. The annotation interface integrates agent profiles, real-time chat functionality, and evaluation metrics. Tutors initiate sessions by diagnosing academic challenges through open-ended questioning, then progressively apply intervention strategies from the academic support guidelines from guide books \citep{gordon2011academic, folsom2015new}, and finally evaluate the performance of student agents.

\subsection{Experts Evaluation Settings}
\label{Annotation Settings}
To ensure the reliability of automated scoring and to facilitate interactive testing for simulated students, we recruited a team of experienced experts with strong backgrounds in education from top universities in China. Each expert had at least one semester of experience as a teaching assistant or academic tutor and was compensated at a rate of 60 RMB/hour. All annotations and interactive tests were conducted in English.

In the \emph{profile score} and \emph{behavior score} annotation tasks, we first presented annotators with the entire interaction between the student agent and the scorer, along with the student agent’s profile information. Following the same instructions used by the LLM scorer, annotators evaluated the consistency of the agent’s profile and behavior. After providing their ratings, annotators were shown the LLM scorer’s results and explanations. They then indicated their level of agreement with the LLM scorer’s assessments.

Subsequent automated filtering and graph‑based score propagation reduced this pool to 17 high‑quality candidate agents. During the Interactive Test stage these 17 agents engaged with academic advising experts in at least 765 dialogue turns overall, producing 1,842 minutes (\~30.7 hours) of simulated advising discourse.

The whole pipeline ensures that the generated student agents not only adhere to logical consistency within their profiles but also exhibit behaviors that are coherent and aligned with their defined attributes, thereby facilitating effective simulations in educational contexts.

\section{Experiments}

We conducted a series of experiments and formulated several research questions to assess the extent to which our evaluation procedure brings simulated-student ranking closer to expert judgments, examine the roles of individual modules, and surface noteworthy phenomena during selection. We provide detailed settings in Appendix~\ref{app:Experimental-Settings}.

\subsection{Experimental Settings}
\subsubsection{Implementation Details}
We set all student agents to use \texttt{GPT-4o} as their backbone and scoring models. All API calls use the default official parameter settings\footnote{https://platform.openai.com/docs/api-reference/introduction}. We release the code and prompts for reproducibility~\footnote{https://anonymous.4open.science/r/student-simulation-0886}.

To substantiate the empirical scope of our study, we randomly instantiated 559 synthetic student profiles that collectively span the demographic and psychometric space defined in Section 3.1. For each profile we conducted four consistency‑scoring dialogues, resulting in 2,236 LLM‑mediated conversations and a cumulative 4,893,486 tokens (8,754 tokens on average per agent).

\begin{table}[t]
    \centering
    \caption{Overall performance of the agent ranking pipeline under various scoring strategies. We compare the initial and propagated phases using profile score ($S_{p}$), behavior score ($S_{b}$), and their average, evaluated with Precision@K, NDCG@K, and Pairwise Accuracy for $K=5$ and $K=|\mathcal{C}|$.}
    \label{tab:overall_performance}
    \begin{tabular}{llcccccc}
    \toprule
    \multirow{2}{*}{\textbf{Score}} & \multirow{2}{*}{\textbf{Phase}} 
        & \multicolumn{3}{c}{\textbf{K=5}} 
        & \multicolumn{3}{c}{\textbf{K=}$|\mathcal{C}|$} \\
    \cmidrule(lr){3-5}\cmidrule(lr){6-8}
    & & \textbf{Prec.} & \textbf{NDCG} & \textbf{PA}
      & \textbf{Prec.} & \textbf{NDCG} & \textbf{PA} \\
    \midrule
    \multirow{2}{*}{$S_{p}$} 
        & Init & 0.0 & 0.0 & 0.0 & 0.0 & 0.0 & 0.0634 \\
        & Prop & 0.2 & 0.3664 & 0.1 & 0.4615 & 0.3695 & 0.0771 \\
    \midrule
    \multirow{2}{*}{$S_{b}$}
        & Init & 0.0 & 0.0787 & 0.3 & 0.1538 & 0.0999 & 0.0559 \\
        & Prop & 0.4 & 0.4200 & 0.2 & 0.4615 & 0.3858 & 0.0777 \\
    \midrule
    \multirow{2}{*}{$\mathrm{Avg}$}
        & Init & 0.0 & 0.0 & 0.0 & 0.2308 & 0.1021 & 0.0670 \\
        & Prop & 0.4 & 0.4200 & 0.2 & 0.4615 & 0.3866 & 0.0775 \\
    \bottomrule
    \end{tabular}
\end{table}

\subsection{Pipeline Performances}
\textbf{RQ1: }\textit{How does the ranking of agents selected through the full pipeline compare to expert rankings in the interactive test?}

To analyze the extent to which the pipeline yields agent rankings aligned with expert preferences, we examine rankings derived from different scoring phases at various stages of the pipeline.
In particular, we consider the profile score, the behavior score, and their average, evaluated at both the initial ($S_{p}, S_b$) and propagated phases ($S^{(K)}_p, S^{(K)}_b$). The rankings are compared using three metrics—Precision@K, NDCG@K, and Pairwise Accuracy—with $K$ set to 5 and to the total number of agents selected ($|\mathcal{C}|$). For all metrics, higher values indicate closer alignment to human preferences. We elaborate on the metrics in Appendix~\ref{app:overall_performance_metrics}.

Table~\ref{tab:overall_performance} reports the ranking performance using different scores throughout the pipeline. We observe that scores computed in the initial scoring stage are less aligned with experts across metrics. In contrast, the average of $S_{p}$ and $S_{b}$ tends to improve the selection quality. Moreover, despite propagation improving top-K precision and NDCG, pairwise accuracy shows only small changes  (< 0.02 absolute improvement) between phases. It indicates that while propagation enhances absolute ranking positions, it preserves most relative agent pairwise relationships established during the two round scoring phase. The marginal improvement implies limited success in correcting initial rankings of pairwise agents through propagation.

\begin{table}[h]
\centering
\small
\caption{Mean Absolute Error comparison of initial and propagated scores compared to human scores.}
\label{tab:MAE_comparison}
\begin{tabular}{c|c|c|c}
\hline
MAE & Init & Propagated & Improv \\ \hline
$\Delta_{profile}$ & 1.007 & 0.6988 & +30.63\% \\
$\Delta_{behavior}$ & 1.6942 & 0.8453 & +50.11\% \\ \hline
\end{tabular}
\end{table}

\begin{wrapfigure}{l}{0.5\linewidth}
  \includegraphics[width=\linewidth]{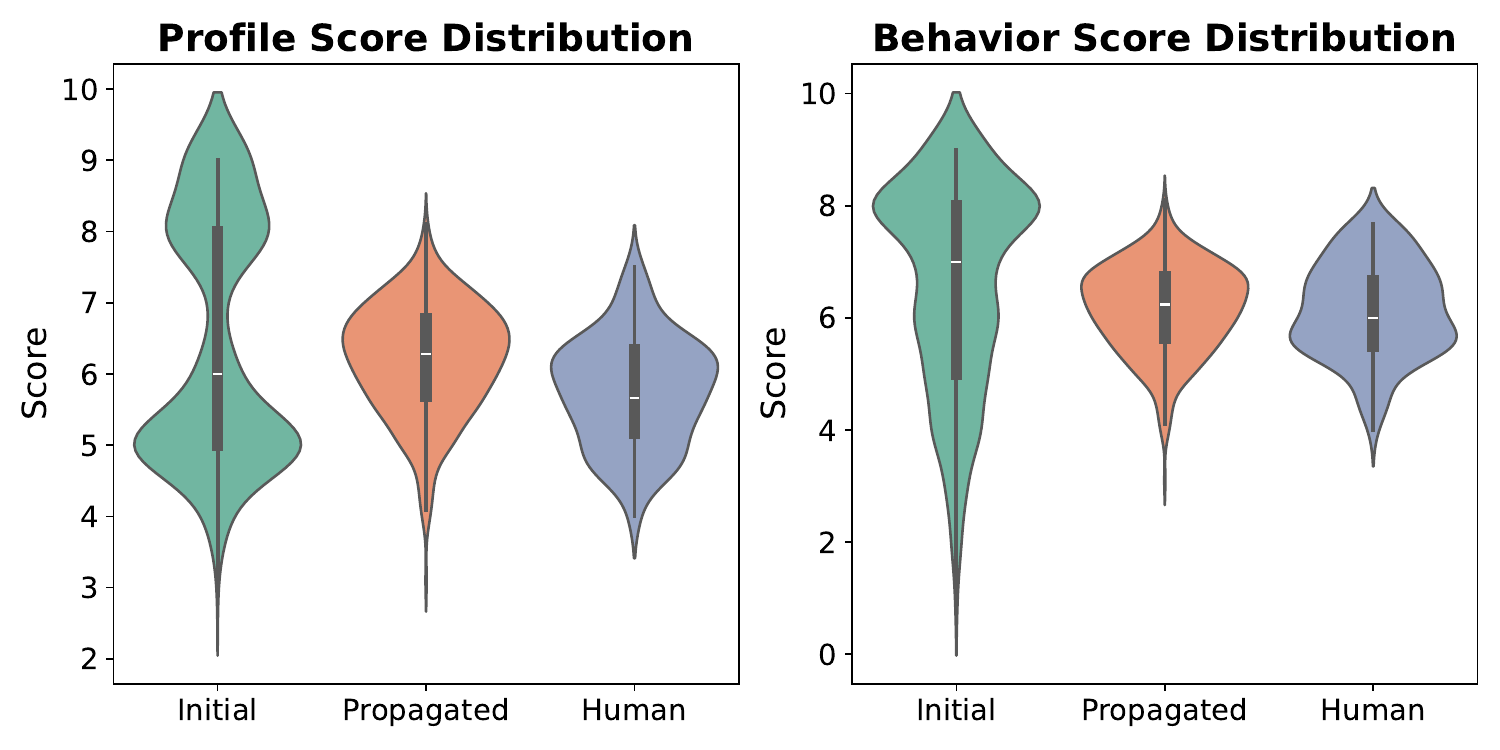}
  \caption{The distributions of the initial (\textcolor[RGB]{72,161,133}{green}), propagated (\textcolor[RGB]{216,110,69}{orange}), and human scores (\textcolor[RGB]{110,128,169}{blue}). Propagated scores are more aligned with that of human scores.}
\label{fig:Initial_vs_Propagated_vs_Human_violin}
\end{wrapfigure}

\subsection{Score Propagation Validation}
\textbf{RQ2: }\textit{To what extent does the score propagation correct the initial scores?}

To examine whether the graph-based score propagation mechanism, we perform comparative analyses between the raw automated scores, the propagated scores, and the average human expert scores. Formally, let $\bar{S}_{\text{expert}}(p_i)$ denote the average score assigned by human experts for the student agent with profile $p_i$. 

As illustrated in Figure \ref{fig:Initial_vs_Propagated_vs_Human_violin}, we present score distributions at the initial and propagated stages alongside human expert scores. After propagation, the distribution more closely tracks that of human annotations. Notably, propagated profile scores exhibit a top-heavy distribution mirroring human annotations, whereas behavior scores remain comparatively bottom-heavy. This pattern suggests that propagation brings profile-consistency assessments closer to human judgments, with smaller effects on behavior-consistency.

We then quantitatively measure the differences between $S^{(K)}(p_i)$ and the average experts' scores $\bar{S}_{\text{expert}}(p_i)$ using Mean Absolute Error (MAE).
The results in Table \ref{tab:MAE_comparison} also demonstrate that the propagated scores have a reduced MAE with expert scores compared to the initial automated scores, thereby validating the effectiveness of the graph-based propagation mechanism.
Specifically, the profile MAE $\Delta_{profile}$ decreases from 1.007 to 0.6988, reflecting a 30.63\% improvement. Similarly, the behavior MAE $\Delta_{behavior}$ reduces from 1.6942 to 0.8453 with a 50.11\% improvement. These observations suggest complementary roles: LLM-based scoring captures explicit profile features, while relational graphs regularize scores via structural similarity. The resulting alignment to expert judgments facilitates downstream analyses of which student profiles and behaviors are simulated more faithfully.

\begin{figure*}[t]
  \centering
  \begin{minipage}{\textwidth}
    \includegraphics[width=\textwidth]{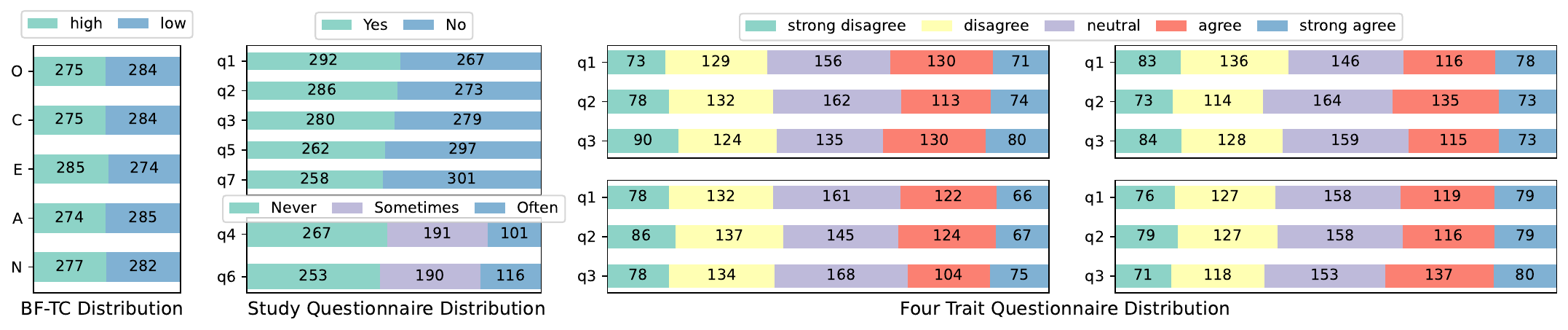}
  \end{minipage}
  
  \begin{minipage}{\textwidth}
    \includegraphics[width=\textwidth]{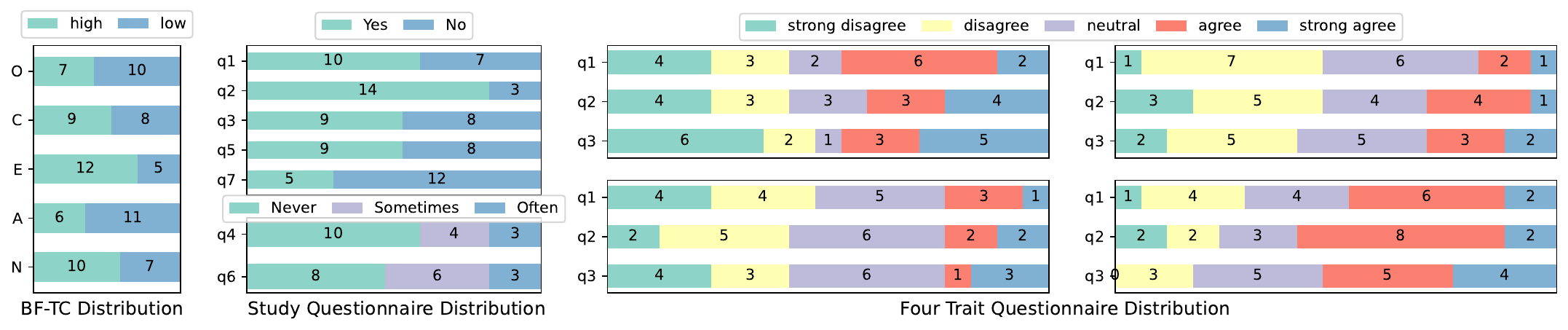}
  \end{minipage}
  \caption{The figure illustrates the distribution of key profile features of student agents before and after the pipeline filtering process. "BF-TC" represents the Big Five personality traits (O: Openness, C: Conscientiousness, E: Extraversion, A: Agreeableness, and N: Neuroticism). Other questionnaire items are labeled by their respective question numbers. The Four Trait Questionnaire subscales are distributed as follows: top-left (goal commitment), bottom-left (motivation), top-right (self-efficacy), and bottom-right (self-regulation).}
  \label{fig:profile_distribution}
\end{figure*}

\subsection{Agent Profile Distribution}
\textbf{RQ3: }\textit{How does the distribution of profile traits shift throughout the pipeline?}

Figure~\ref{fig:profile_distribution} illustrates the distribution of key profile features of student agents before and after the pipeline filtering process. The initial distribution appears uniform, with minor differences across features. However, most feature distributions exhibit noticeable shifts in the final selected agents.

Among the Big Five personality traits, agents with low extraversion and high agreeableness are more prevalent in the final selection. It is because high extraversion often indicates strong energy and positivity, which may not align well with typical characteristics of students experiencing learning difficulties. In contrast, high agreeableness is more frequently observed, as agents tend to exhibit affirmative responses, making agents with high agreeableness more likely to receive higher alignment scores.
For the study questionnaire, q2 (\textit{Do you have trouble taking tests or completing assignments?}) shows a significantly higher proportion of "Yes" responses in the selected agents. It suggests that LLM-based student simulations better capture students with such difficulties, aligning with real-world academic advising practices where these challenges are common.
Notably, in the four trait questionnaire, the proportion of agents with strong self-efficacy and self-regulation increases substantially after filtering. It is because agents inherently display high self-efficacy and structured planning abilities in conversations, even though these traits may not typically align with students facing learning difficulties. Nevertheless, such agents still achieve high scores in behavior consistency evaluations.

\begin{figure}[t]
  \includegraphics[width=\columnwidth]{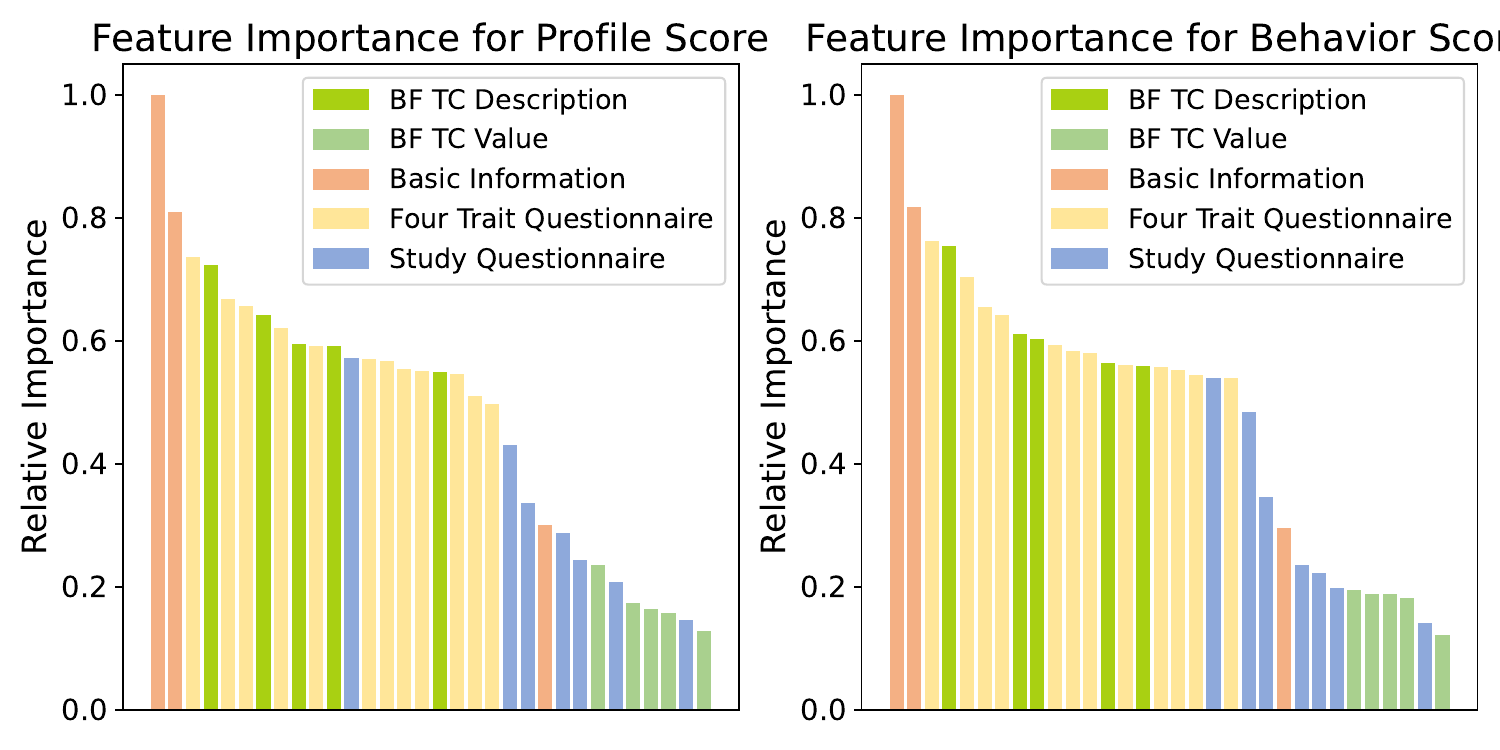}
  \caption{Relative importance ranking of traits in profile, with those under the same level categorized by colors.}
  \label{fig:feature_importance}
\end{figure}

\subsection{Influence of Features in Profile}
\textbf{RQ4: }\textit{How important are the traits to final scores?}

To investigate the influence of the features in the agent profile on the final profile and behavior scores, we performed a hot encoding of the features of the profile and used a Random Forest model \citep{breiman2001random} to fit the propagated profile score and behavior score, where the size of test sets is 20\% and the number of trees is 100. We then ranked the features based on their relative importance\citep{archer2008empirical}, where the relative importance of each feature is calculated by dividing its importance score by the highest among all. 

As shown in Figure \ref{fig:feature_importance}, the profile features are categorized into five groups, with \textit{BF TC} containing both high-low value-based features and their corresponding descriptive features.
\textbf{Age} and \textbf{MBTI} personality type from the \textit{Basic Information} category have a significant impact on the scores. It indicates that both profile and behavior scorer pay more attention to these two traits when scoring the agents. Additionally, the overall influence of the \textit{Four Trait Questionnaire} is relatively high, with similar importance observed for the descriptive features of \textit{BF TC}. Compared to value-based features, descriptive features provide more detailed information to the scorer, aiding in assessing of the agent's consistency.
Notably, even the most important feature \textbf{Age} has a relatively low importance score of 0.0651 and 0.0644 for the profile and behavior scores, respectively. It suggests that individual features have limited influence on the final scores, and the interaction between multiple features plays a more critical and subtle role in determining the outcomes.

\subsection{Case Study}
\textbf{RQ5: }\textit{Which trait does the LLM prioritize or overlook when evaluating the quality of simulations?}

\begin{figure*}[h]
  \centering
  \begin{minipage}{\textwidth}
    \includegraphics[width=\textwidth]{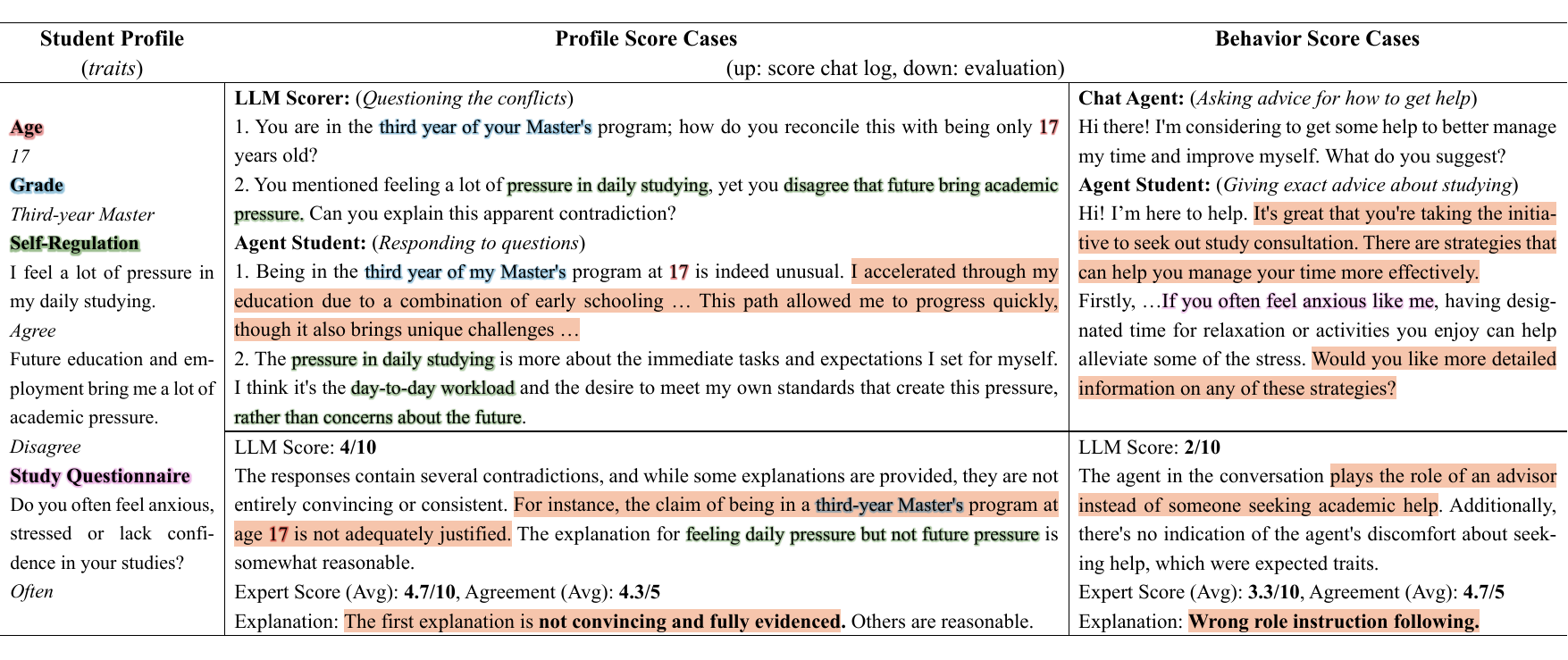}
  \end{minipage}
  \begin{minipage}{\textwidth}
    \includegraphics[width=\textwidth]{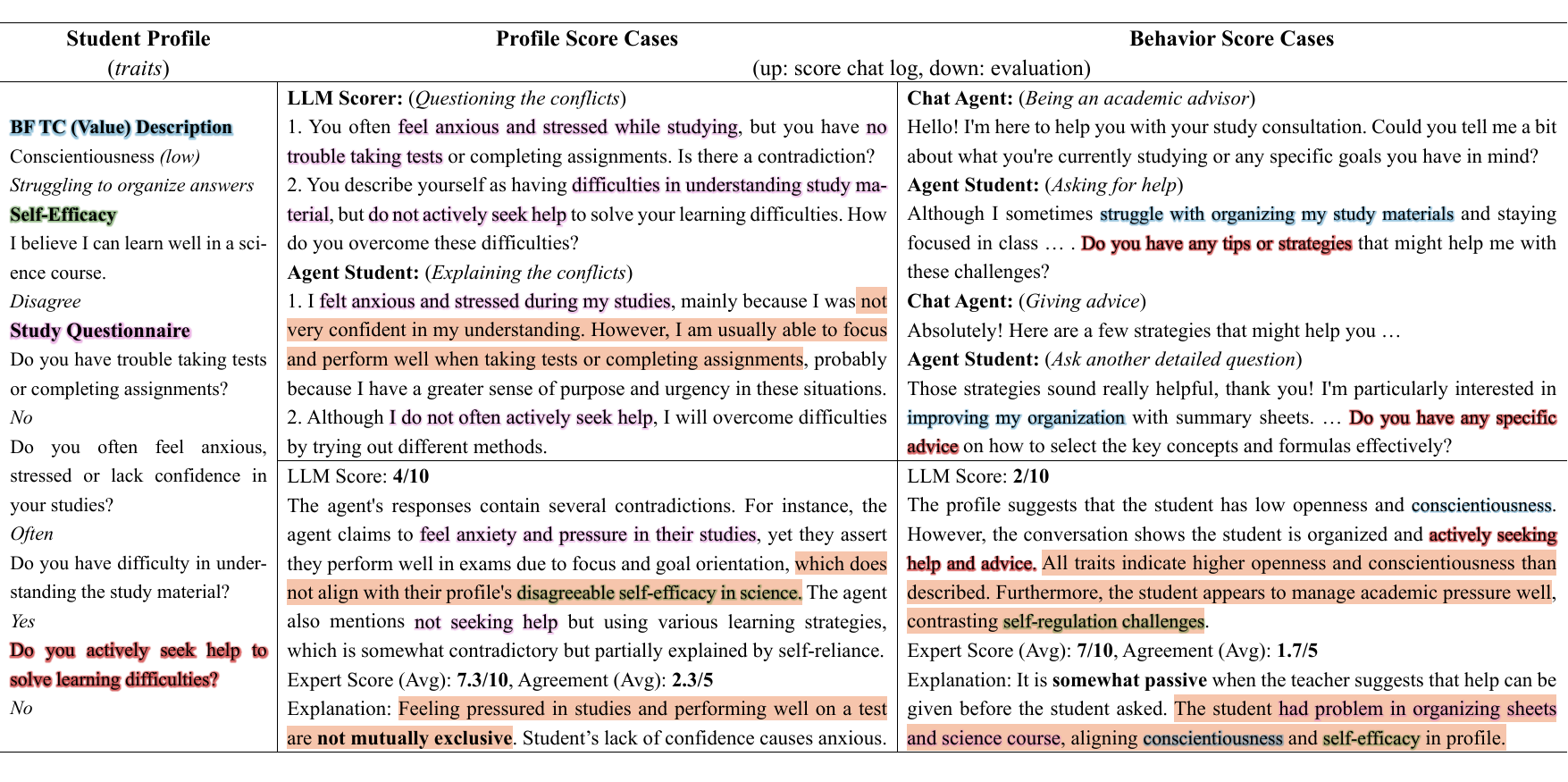}
  \end{minipage}
  \caption{Examples of LLM scoring cases for agent profile and behavior: good case (top) and bad case (bottom). Colored text highlights corresponding features in the profile.}
  \label{fig:cases}
\end{figure*}

We showcase LLM scoring examples in Figure~\ref{fig:cases}, comparing a \textit{good case} and a \textit{bad case} in profile and behavior evaluation, accompanied by expert scores, agreement levels, and explanations. In the \textbf{good case} (top), the questioning agent $Q$ detects an inconsistency in the student's profile—17 years old while enrolled as a third-year Master's student. The student agent fails to provide a compelling justification for this contradiction. Additionally, the LLM accurately assesses the student agent's self-reported academic pressure, where the reasoning behind the discrepancy between daily and long-term pressure is somewhat reasonable. Regarding behavior evaluation, the scorer correctly identifies that the student agent violates the role of being a student who need academic advising. It demonstrates the model's sensitivity to instruction following and reveals its occasional struggle with adherence to long instructions.

In contrast, the \textbf{bad case} (bottom) illustrates a scenario where the LLM's evaluation deviates significantly from human judgment. The agent's profile presents conflicting traits, such as experiencing stress while reporting no difficulty in tests and assignments. The LLM questions these inconsistencies but assigns a relatively low score due to its strict conflict detection. However, expert annotations suggest that such contradictions are not inherently erroneous, but reflect common real-world cases where students manage stress effectively despite academic challenges. Furthermore, the LLM scorer underperforms behavior evaluation by misinterpreting the agent's conscientiousness and self-efficacy, leading to lower agreement scores with human experts. The results indicate that while LLM scorers is adept at identifying explicit conflicts, it may over-penalize cases where nuanced human explanations exist.

These findings suggest that the LLM tends to overemphasize explicit contradictions while occasionally overlooking the complexity of real-world student behaviors, reflected in certain cases' lower agreement scores with human experts. Future improvements should focus on enhancing the model's reasoning consistency and refining its instruction-following capabilities to align more closely with expert evaluations.

\section{Conclusion}
We propose a pipeline for automatically generating and filtering high-quality simulated student agents to improve the authenticity of learning difficulty simulations. Our approach contains a two-round automated scoring phase, validated by human experts to ensure the profile consistency and behavioral alignment of agents. We then employ a score propagation module to enhance consistency across the student graph. Experimental results show that our method effectively identifies high-quality agents, advancing the use of simulated students for broader educational applications.

\section*{Limitations}
Our work explores the potential of automatically generating and filtering high-quality simulated student agents to enhance the authenticity of learning difficulty simulations. However, researchers and practitioners must consider several limitations and risks.
First, our approach focuses on prompt-based student simulation, particularly extending the scope to metacognition. We do not explore more complex agent simulation frameworks like workflow-based or retrieval-augmented generation (RAG) simulations. These approaches will be explored in future work to improve simulation fidelity further.
Second, some studies evaluate LLM-based human simulation by Turing tests \citep{aher2023using}, measuring the model's success rate in mimicking human-like responses. Such evaluations emphasize linguistic style. Our work prioritizes metacognitive abilities and learning limitations in simulation rather than stylistic authenticity. We explicitly addressed this distinction during expert annotation to ensure the evaluation remains aligned with our research objectives.
Third, our approach assumes that structured profiles can effectively capture student attributes and learning difficulties. However, real-world learning behaviors are highly dynamic and context-dependent. Future work should explore adaptive modeling techniques that better reflect the evolving nature of student cognition and engagement.

\section{Acknowledgements}
This work was supported by the SMP-Zhipu.AI Large Model Cross-Disciplinary Fund (Grant No. 20240211) and the National Natural Science Foundation of China (Grant No. 62407027).

\section{Author Contributions}
Haoxuan Li was responsible for the writing and revision of this paper, experimental design, and implementation; Jifan Yu, Xin Cong were responsible for advising the experimental design and writing the paper; Yang Dang was responsible for the experimental design; Daniel Zhang-Li, Lu Mi, Yisi Zhan, Huiqin Lin, Zhiyuan Liu were responsible for advising the writing and providing guidance on the research route of the paper. All authors approved the version to be published and agree to be accountable for all aspects of the work in ensuring that questions related to the accuracy or integrity of any part of the work are appropriately investigated and resolved.

\appendix

\section{Ethical Considerations}
Simulated student agents are generated based on training data, which may introduce biases, stereotypes, or unintended inaccuracies in student representations. These biases could affect the fairness and applicability of educational interventions. Researchers must remain vigilant in mitigating such risks through continuous evaluation, bias detection, and responsible deployment of simulated agents.

\section{Experimental Settings}
\label{app:Experimental-Settings}




\subsection{Pipeline Performance Metrics}
\label{app:overall_performance_metrics}

The evaluation of ranking performance relies on three widely used metrics: Precision@K, Normalized Discounted Cumulative Gain at K (NDCG@K), and Pairwise Accuracy. These metrics capture different aspects of ranking quality, ensuring a comprehensive assessment of the proposed approach.

Precision@K measures the proportion of relevant agents among the top $K$ positions in the ranking. Formally, it is computed as: 
\begin{equation} 
\text{Precision@K} = \frac{|{r_i \in R_K \cap G}|}{K}, \end{equation} 
where $R_K$ denotes the set of top $K$ agents according to the pipeline ranking, and $G$ represents the set of agents deemed relevant based on expert annotations. A higher Precision@K indicates a greater concentration of relevant agents in the top-ranked positions.

NDCG@K extends the evaluation by incorporating graded relevance, ensuring that highly relevant agents are given more weight when positioned near the top of the ranking. It is defined as: 
\begin{equation} 
\text{NDCG@K} = \frac{1}{\text{IDCG@K}} \sum_{i=1}^{K} \frac{2^{\text{rel}(r_i)} - 1}{\log_2 (i+1)}, \end{equation} 
where $\text{rel}(r_i)$ denotes the relevance score of agent $r_i$, and $\text{IDCG@K}$ represents the maximum possible DCG@K value obtained from an ideal ranking where all agents are perfectly ordered by their relevance scores. The logarithmic discount factor penalizes lower-ranked relevant agents, emphasizing the importance of placing highly relevant agents at the top of the ranking. Higher NDCG@K values indicate rankings that more closely align with the ideal ordering.

Pairwise Accuracy evaluates ranking correctness at the level of agent pairs by comparing the relative ordering produced by the system against the ground-truth order determined by expert annotations. It is given by: \begin{equation} 
\text{Pair Acc.} = \frac{\sum_{(i,j)\in P} \mathbf{1}\big[(r_i - r_j)(g_i - g_j) > 0\big]}{|P|}, \end{equation} 
where $P$ represents the set of all possible agent pairs, and $\mathbf{1}(\cdot)$ is an indicator function that returns one if the system correctly preserves the relative ordering of a pair and zero otherwise. A higher pairwise accuracy indicates better agreement between the system and expert-annotated rankings.

Together, these metrics provide a comprehensive evaluation of ranking quality. Precision@K focuses on binary relevance at a fixed cutoff, NDCG@K captures position-sensitive graded relevance, and Pairwise Accuracy measures local ranking consistency.

\section{Profile Settings}
\label{app:profile}

\begin{figure}[t]
  \includegraphics[width=\columnwidth]{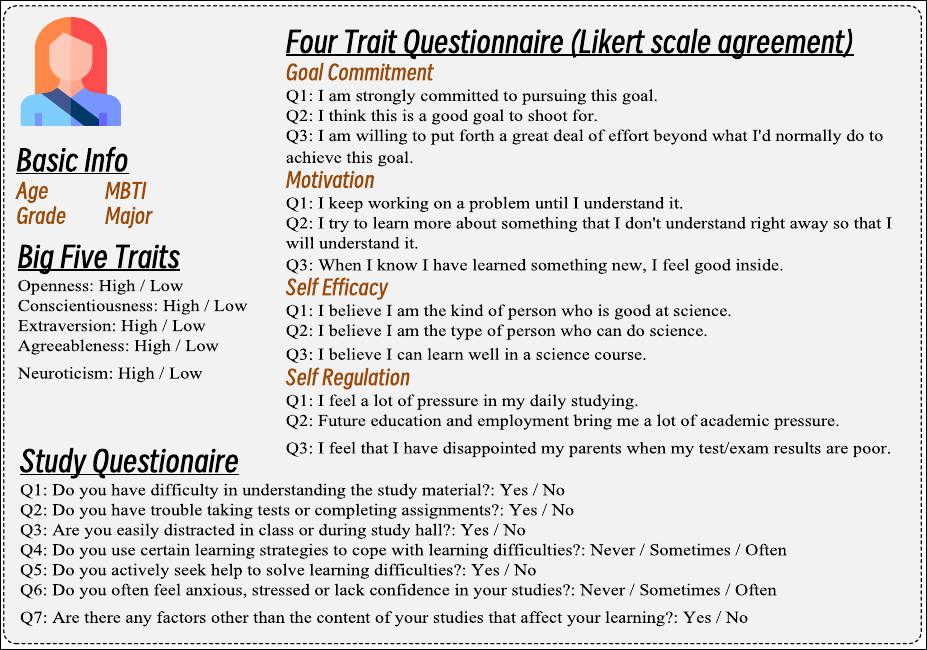}
  \caption{Student profile includes demographic information and other studying-related questionnaires.}
  \label{app:student_profile}
\end{figure}

The virtual student profiles are constructed through four demographic attributes and two psychological assessments, as visualized in Figure~\ref{app:student_profile}. Demographic components include:

\begin{itemize}
[
  itemsep=0pt,      
  parsep=0pt,       
  topsep=0pt,       
]
    \item \textbf{Basic Attributes}: Gender, age (17-28 years covering typical undergraduate/graduate student ranges), academic major (62 most prevalent disciplines across Science, Engineering, Social Science and Arts), and academic standing (freshman to third-year master).
    \item \textbf{Psychological Profiles}: MBTI (16 personality types) and Big Five trait compositions with the corresponding descriptions.
\end{itemize}

Four multi-dimensional questionnaires capture learning characteristics:
\begin{itemize}
[
  itemsep=0pt,      
  parsep=0pt,       
  topsep=0pt,       
]
    \item \textbf{Learning Traits}: 12 Likert-scale items (1-5 agreement) measuring self-regulation, motivation, engagement, and information processing styles. Each dimension contains three semantically equivalent questions for response consistency verification.
    \item \textbf{Academic Challenges}: 7 common learning obstacles across four domains: cognitive (e.g., material understanding), metacognitive (e.g., learning strategies), social (e.g., seeking for help), and affective (e.g., stress management). 
\end{itemize}

\newpage
\bibliographystyle{elsarticle-num} 
\bibliography{sample-base}





\end{document}